\documentclass[11pt]{amsart}
\usepackage[margin=1in]{geometry}		
\usepackage[parfill]{parskip}			
\usepackage{graphicx}
\usepackage{amssymb}
\usepackage{epstopdf}
\title{The Dirac equation without zitterbewegung}
\author{Michael B. Heaney\\3182 Stelling Drive\\Palo Alto, CA 94303\\mheaney@alum.mit.edu}
\date{9 October 2014}				
\begin{document}
\maketitle
\begin{abstract}
This paper presents a relativistic symmetrical interpretation of the Dirac equation in 1+1 dimensions which predicts no zitterbewegung for a free spin-$\frac{1}{2}$ particle. This could resolve the longstanding puzzle of zitterbewegung in relativistic quantum mechanics, and help clarify the physical meaning of the zitterbewegung currently under investigation in many-particle systems. Together with an earlier paper describing a relativistic symmetrical interpretation of the Klein-Gordon equation \cite{Heaney1}, this new interpretation resolves some inconsistencies in the conventional interpretations of both equations. This new interpretation also makes several experimental predictions that differ from the conventional interpretation of the Dirac equation.
\end{abstract}
\section{Introduction}
The Dirac Equation (DE) is the most important equation of quantum mechanics. Despite this importance, the DE is still not completely understood. For example, in 1930 Schr\"odinger discovered that the Conventional Interpretation (CI) of the DE predicts a rapid oscillating motion for a free spin-$\frac{1}{2}$ particle in vacuum, naming it zitterbewegung \cite{Schroedinger}. Zitterbewegung of a free electron has a frequency of order $10^{21}$ Hz and an amplitude of order $10^{-13}$ m, which is beyond measurement with current technology. But why would a single electron in free space, with no forces acting on it, spontaneously move back and forth at the speed of light? This seems unphysical. Some physicists believe zitterbewegung is not a real physical phenomenon, but just an artifact of an incorrect single-particle interpretation of the DE \cite{Itzykson}. Other physicists believe zitterbewegung is a real physical phenomenon, with experimentally observable effects \cite{Sakurai}. Zitterbewegung of a free particle is an unresolved puzzle in the foundations of the CI of quantum mechanics. 

Zitterbewegung has recently become a topic of intensive investigation: \textquotedblleft From an obscure, perplexing and somewhat marginal effect that would probably never be observed, zitterbewegung has grown into a universal, almost ubiquitous phenomenon that was experimentally simulated in its quantum form and directly observed in its classical version. Our article summarizes the first five years of the intensive development which can be characterized as the \textquoteleft Sturm und Drang' period..." \cite{Zawadzki1}. This intensive investigation includes new theoretical work attempting to resolve the puzzle of zitterbewegung of a free spin-$\frac{1}{2}$ particle obeying the DE.

The CI of quantum mechanics, which predicts zitterbewegung, is favored by less than half of physicists working on the foundations of quantum mechanics \cite{Schlosshauer}. An alternative interpretation, the Two-State Vector Formalism (TSVF), and the related discovery of \textquotedblleft weak measurements," have become major areas of research over the past decade \cite{Struppa}. The TSVF is one example of a Symmetrical Interpretation (SI) of quantum mechanics. The basic idea of all SI's is to restore time symmetry to quantum mechanics. SI's of quantum mechanics have a long history, dating back at least to a 1921 paper by Schottky \cite{Schottky}. Many different types of SI's have been developed over the past century [7-33]. The similarities and differences between the TSVF, the interpretation presented in this paper,  and other types of SI were described in an earlier paper \cite{Heaney1}.

We now give a conceptual motivation for a Relativistic Symmetrical Interpretation (RSI) of quantum mechanics. The CI postulates that the transition of a particle from an initial location $(x_i,t_i)$ to a final location $(x_f,t_f)$ takes place via a retarded wavefunction which starts localized around position $x_i$ at time $t_i$, then travels forwards in time towards position $x_f$ while also becoming delocalized, then collapses into a different wavefunction localized around position $x_f$ at time $t_f$. However, all relativistic wave equations have two distinct types of solutions: retarded waves and advanced waves. The retarded solutions can be interpreted as waves moving forwards in time, and the advanced solutions can be interpreted as waves moving backwards in time, without violating the special theory of relativity \cite{Wheeler}. Because of this, the transition of a particle from $(x_i,t_i)$ to $(x_f,t_f)$ can be explained equally well by an advanced interpretation: an advanced wavefunction starts localized around position $x_f$ at time $t_f$, then travels backwards in time towards position $x_i$ while also becoming delocalized, then collapses into a different wavefunction localized around position $x_i$ at time $t_i$. A more detailed and quantitative comparison of these two possible interpretations has been given elsewhere \cite{Heaney2}. These two alternatives are not classically distinguishable, suggesting we should not add their amplitudes to get the total amplitude of the transition. Instead, if we generalize the classical concept of sequential transitions to include a quantum particle traveling from $(x_i,t_i)$ to $(x_f,t_f)$ and then traveling back to $(x_i,t_i)$, the total amplitude for the transition will be the product of the two amplitudes. The product of these two amplitudes is interpreted as the RSI transition amplitude. This gives the same experimental predictions for transitions as the CI transition amplitude, but with a significantly different interpretation. 
\section{The Conventional Interpretation of the Dirac Equation}
We will first describe how the CI of the DE predicts zitterbewegung for a free spin-$\frac{1}{2}$ particle in a vacuum. The CI implicitly assumes that quantum mechanics is a theory which describes an isolated, individual physical system, such as a single electron in free space. Given the CI founders past experience exclusively with classical mechanics, this must have seemed too obvious to postulate. The CI wavefunction postulate assumes such a system is maximally described by a wavefunction $\psi(x,t)$, together with specified initial conditions. The CI evolution postulate assumes the wavefunction of a free, spin-$\frac{1}{2}$ particle in 1+1 dimensions will evolve continuously and deterministically forwards in time from the initial conditions according to the relativistic DE:

\begin{equation}
\frac{1}{c}\sigma_0\frac{\partial\psi}{\partial t} + \sigma_x\frac{\partial\psi}{\partial x} + \frac{imc}{\hbar}\sigma_0\psi=0,
\label{ }
\end{equation}
where
\begin{equation}
\sigma_0 =
\left(
\begin{array}{ccc}
1 & 0\\
0 & 1\\
\end{array}
\right),
\label{ }
\end{equation}
\begin{equation}
\sigma_x =
\left(
\begin{array}{ccc}
0 & 1\\
1 & 0\\
\end{array}
\right),
\label{ }
\end{equation}
\begin{equation}
\psi =
\left(
\begin{array}{c}
\psi_1 \\
\psi_2 \\
\end{array}
\right),
\label{ }
\end{equation}
$c$ is the speed of light, $m$ is the mass of the particle, and $2\pi\hbar$ is Planck's constant. The CI founders viewed the retarded wavefunction equation of a particle, with specified initial conditions for the wavefunction, as the quantum equivalent of Newton's second law for a classical particle, with specified initial conditions for the particle's position and momentum. The alternative possibility of using the advanced wavefunction equation with specified final conditions was not even considered.

The Hermitian conjugate of the DE is also a valid wavefunction equation:

\begin{equation}
\frac{1}{c}\frac{\partial \psi^\dagger}{\partial t}\sigma_0 + \frac{\partial \psi^\dagger}{\partial x}\sigma_x - \frac{imc}{\hbar}\psi^\dagger\sigma_0=0.
\label{ }
\end{equation}

If we multiply the DE on the left by $\psi^\dagger$, multiply the Hermitian conjugate of the DE on the right by $\psi$, and add the two resulting equations, we get a CI local conservation law:

\begin{equation}
\frac{\partial\rho}{\partial t}+\nabla\cdot j=0,
\label{ }
\end{equation}
where
\begin{equation}
\rho(x,t)\equiv\psi^\dagger\psi,
\label{ }
\end{equation}
and 
\begin{equation}
j(x,t)\equiv c\psi^\dagger\sigma_x\psi.
\label{ }
\end{equation}

The CI interprets $\rho(x,t)$ as the probability density for finding the particle at position $x$ at time $t$, and $j(x,t)$ as the probability density current. The fact that $\rho(x,t)$ is always real and positive is taken as a confirmation of this interpretation. 

If $\psi(x,t)$ is normalized and goes to zero at $x=\pm\infty$, we can integrate the CI local conservation law over all space to get a CI global conservation law:

\begin{equation}
\int_{-\infty}^{+\infty}\rho(x,t)dx = 1,
\label{ }
\end{equation}

which means the probability of finding the particle somewhere in space is conserved and equal to one. 

Figure 1(a) shows the CI probability density at the initial time $t_i\equiv 0$ for a normalized gaussian wavefunction given by:

\begin{equation}
\psi(x,0) \equiv
\left(
\frac{1}{32\pi}
\right)^{1/4}e^{-\frac{x^2}{16}}
\left(
\begin{array}{c}
1 \\
1 \\
\end{array}
\right),
\label{ }
\end{equation}

where we use natural units and the standard deviation is 2 \cite{Thaller}. This wavefunction has both positive energy and negative energy components, as required by the CI to obtain complete solutions of the DE. It is the interference between these two components which causes zitterbewegung. Figures 1(a-c) show the space-time evolution of the probability density according to the DE. The CI probability density $\rho(x,t)$ shows strange distortions as it evolves. In particular, the mean position of the probability density oscillates rapidly with time. These rapid oscillations are the zitterbewegung \cite{Schroedinger}. Figure 2 shows the first cycle of this zitterbewegung. The CI probability density $\rho(x,t)$ also becomes more delocalized with time, up until the measurement at the final time $t_f\equiv 40$. 

 Let us assume that a measurement of the location of the particle at the final time $t_f\equiv 40$ gives the same wavefunction and probability density as at $t_i\equiv 0$, as shown in Fig. 1(d). The CI measurement postulate assumes the transition amplitude $A$ for a particle having the initial wavefunction $\psi(x,t_f-\delta t)$ to be found having the final wavefunction $\phi(x,t_f+\delta t)$ is:

\begin{equation}
A\equiv\int_{-\infty}^{+\infty}\phi^\dagger(x,t_f+\delta t) \psi(x,t_f-\delta t)dx.
\label{ }
\end{equation}

For the transition from Fig. 1(c) to Fig. 1(d), numerical calculations give $A = -0.584 -  0.010 i$. The CI measurement postulate also assumes the transition probability $P$ for such a transition is:

\begin{equation}
P\equiv A^\ast A.
\label{ }
\end{equation}

For the transition from Fig. 1(c) to Fig. 1(d), numerical calculations give $P=0.341$.

The CI collapse postulate assumes that, upon measurement at $t_f$, the wavefunction $\psi(x,t_f-\delta t)$, and therefore the CI probability density of Fig. 1(c), undergoes an instantaneous, indeterministic, and time-asymmetric collapse into the CI probability density of Fig. 1(d). Conservation laws require this collapse to be instantaneous in all reference frames. This collapse does not obey the DE, nor any other known equation. The CI reinterpretation of the DE as a quantum field equation gives no further explanation of how this wavefunction collapse occurs \cite{Nikolic}. This collapse introduces an arrow of time into the CI of quantum mechanics, destroying the time symmetry of the DE. The CI postulates that the new wavefunction $\phi(x,t_f+\delta t)$ of Fig. 1(d) will then evolve forwards in time according to the DE.
 \begin{figure}[htbp]
\begin{center}
\includegraphics[width=6.5in]{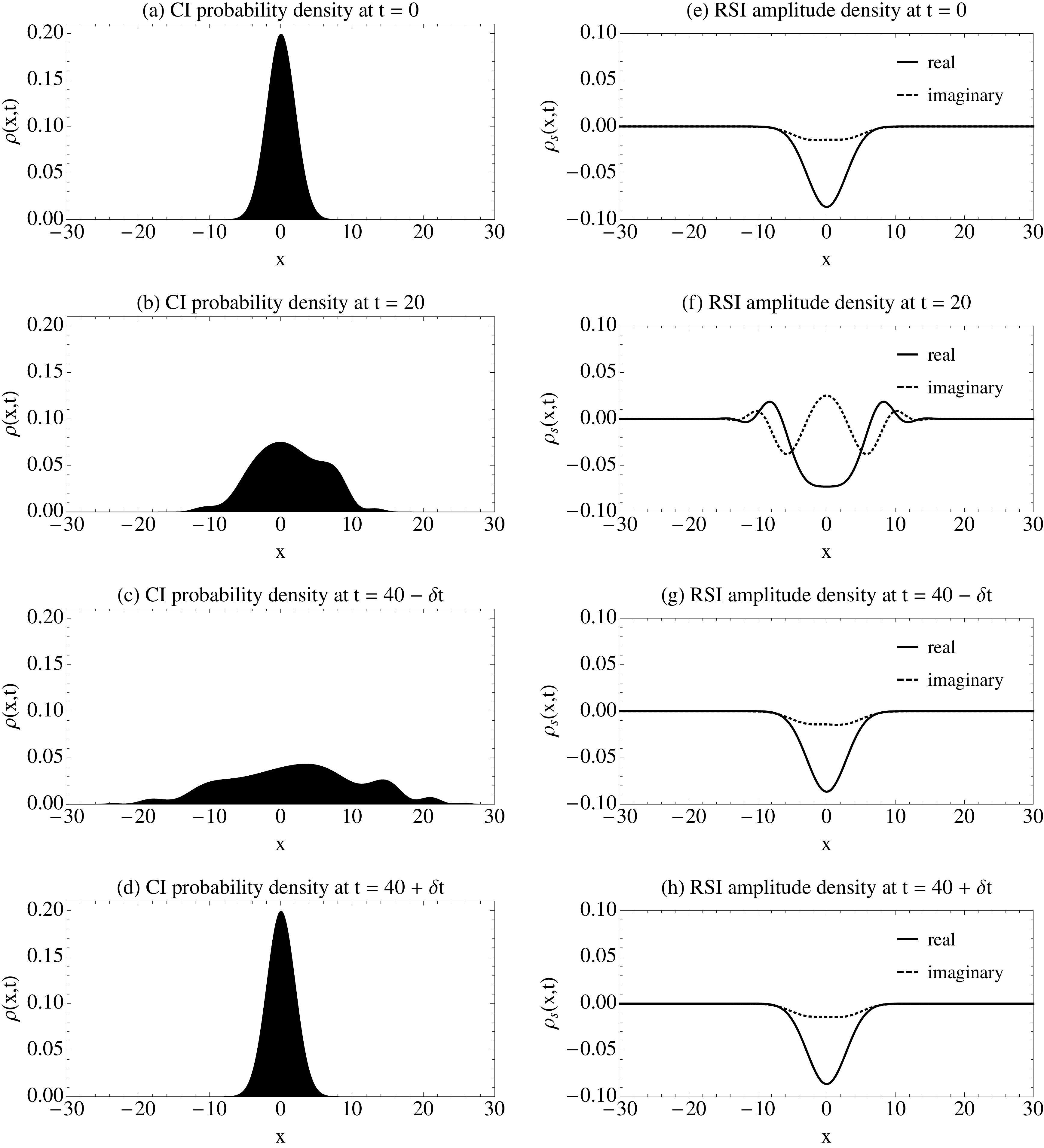}
\caption{\textbf{(a-d):} The Conventional Interpretation (CI) of the Dirac equation for a free particle, evolving from a gaussian probability density at $t=0$ to an identical probability density at $t=40$. The CI probability density $\rho\equiv\psi^\dagger\psi$ shows strange distortions as it evolves. Upon measurement at $t=40$, the CI postulates that the wavefunction of (c) collapses instantaneously and irreversibly to the wavefunction of (d). \textbf{(e-h):} The Relativistic Symmetrical Interpretation (RSI) of the Dirac equation for a similar transition. The real and imaginary parts of the RSI transition amplitude density $\rho_s\equiv\phi_+^\dagger\psi_+$ evolve smoothly and continuously at all times, with no strange distortions. The RSI postulates no wavefunction or transition amplitude collapse at any time. The negative energy RSI transition amplitude density $\phi_-^\dagger\psi_-$ gives the same results but with reversed phase, suggesting interpretation as the antiparticle.}
\label{default2}
\end{center}
\end{figure}
 \begin{figure}[htbp]
\begin{center}
\includegraphics[width=6.5in]{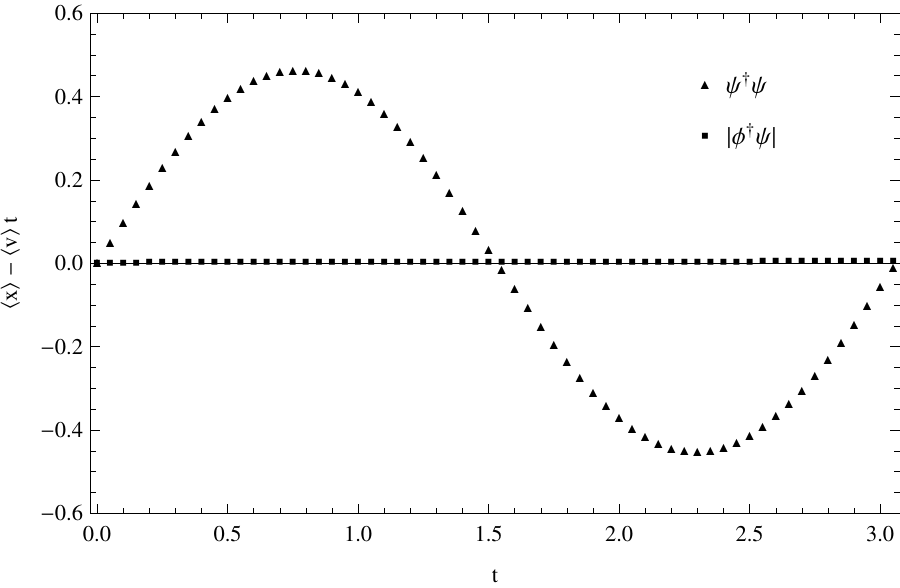}
\caption{The mean positions $\langle x \rangle$ of the CI probability density $\rho\equiv\psi^\dagger\psi$ and of the absolute value of the positive energy RSI amplitude density $|\rho_s|\equiv|\phi_+^\dagger\psi_+|$, with the drift motions $\langle v \rangle t$ subtracted out. The CI probability density shows the first cycle of zitterbewegung, while the RSI amplitude density shows no significant zitterbewegung. The small $(\pm 0.006)$ oscillations visible in the RSI amplitude density are present in the wavefunctions $\psi_+$ and $\phi_+^\dagger$, and decrease as the numerical accuracy of the calculations is increased, suggesting they are caused by numerical imprecision. The negative energy RSI transition amplitude density $\phi_-^\dagger\psi_-$ gives the same results.}
\label{default2}
\end{center}
\end{figure}
\section{The Relativistic Symmetrical Interpretation of the Dirac Equation}
We will now develop a Relativistic Symmetrical Interpretation (RSI) of the DE, and show how it predicts no zitterbewegung for a free spin-$\frac{1}{2}$ particle in a vacuum. Dirac showed that all of the experimental predictions of the CI can be formulated in terms of transition probabilities \cite{Dirac}. The RSI inverts this observation to explicitly assume that quantum mechanics is a theory which describes transitions, such as the transition of a single electron between two localized states in free space. The RSI evolution postulate assumes such a transition is maximally described by a positive energy retarded wavefunction $\psi_+(x,t)$ satisfying the initial conditions, a negative energy advanced wavefunction $\phi_+^\dagger(x,t)$ satisfying the final conditions (the Hermitian conjugation changes the positive energy wavefunction $\phi_+(x,t)$ into a negative energy wavefunction), and the algebraic product of these wavefunctions, where $\psi_+(x,t)$ obeys the DE:

\begin{equation}
\frac{1}{c}\sigma_0\frac{\partial\psi_+}{\partial t} + \sigma_x\frac{\partial\psi_+}{\partial x} + \frac{imc}{\hbar}\sigma_0\psi_+=0,
\label{ }
\end{equation}

and $\phi_+^\dagger(x,t)$ obeys the Hermitian conjugate of the DE:

\begin{equation}
\frac{1}{c}\frac{\partial \phi_+^\dagger}{\partial t}\sigma_0 + \frac{\partial \phi_+^\dagger}{\partial x}\sigma_x - \frac{imc}{\hbar}\phi_+^\dagger\sigma_0=0.
\label{ }
\end{equation}

The RSI treats the wavefunctions as intermediate steps to the symmetrical transition amplitude density, which is the primary object of interest. The negative energy parts of these wavefunctions, $\psi_-$ and $\phi_-^\dagger$, will be seen to describe an antiparticle transition amplitude. The RSI postulate that the amplitude density is a product of purely positive or purely negative energy wavefunctions guarantees that zitterbewegung cannot occur. Note that this postulate was based on an earlier RSI of the KGE \cite{Heaney1} and the conceptual reasoning about time symmetry given in the introduction, and is not an ad hoc postulate to eliminate zitterbewegung.

If we multiply Eq. 13 on the left by $\phi_+^\dagger$, multiply Eq. 14 on the right by $\psi_+$, and add the two resulting equations, we get a RSI local conservation law: 

\begin{equation}
\frac{\partial\rho_s}{\partial t}+\nabla\cdot j_s=0,
\label{ }
\end{equation}
where
\begin{equation}
\rho_s(x,t)\equiv\phi_+^\dagger\psi_+,
\label{ }
\end{equation}
and 
\begin{equation}
j_s(x,t)\equiv c\phi_+^\dagger\sigma_x\psi_+.
\label{ }
\end{equation}

The RSI interprets $\rho_s(x,t)$ as the transition amplitude density, and $j_s(x,t)$ as the transition amplitude density current. These are generally complex functions, and cannot be interpreted as probability densities. The RSI transition amplitude $A_s$ is defined as the integral over all space of the RSI transition amplitude density:

\begin{equation}
A_s\equiv \int_{-\infty}^{+\infty}\phi_+^\dagger(x,t)\psi_+(x,t)dx.
\label{ }
\end{equation}

$A_s$ is a complex constant, independent of the time $t$ and the position $x$. $A_s$ is the amplitude that a particle found in the initial state $\psi_+$ at time $t_i$ will later be found in the final state $\phi_+$ at time $t_f$. The RSI transition probability $P_s$ is defined as $P_s\equiv A_s^\ast A_s$. $P_s$ is the probability that a particle found in the initial state $\psi_+$ at time $t_i$ will later be found in the final state $\phi_+$ at time $t_f$. 

Assuming the same initial ($\psi_+$) and final ($\phi_+$) wavefunctions, the RSI gives the same results for the transition amplitude and probability as the CI, but with a significantly different interpretation: The CI postulates that the wavefunction $\psi_+(x,t)$ evolves continuously and smoothly from time $t_i$ to time $t_f$, then at time $t_f$ collapses instantaneously into the wavefunction $\phi_+(x,t_f)$, similar to Fig. 1(a-d). This is why the CI transition amplitude of Eq. (11) is evaluated at the time of collapse $t_f$. The RSI postulates that $\rho_s(x,t)$ evolves continuously and smoothly from time $t_i$ to times beyond $t_f$ (assuming ideal measurements) as shown in Fig. 1(e-h). This is why the RSI transition amplitude of Eq.(18) is not evaluated at any particular time. The CI has a built-in arrow of time, while the RSI has no arrow of time. 

Let us assume $\psi_+(x,t)$ is the positive energy part of $\psi(x,t<t_f)$, and $\phi_+^\dagger(x,40)$ is the positive energy part of $\phi^\dagger(x,t>t_f)$. Figures 1(e,f,g,h) show the space-time evolution of the real and imaginary parts of the RSI transition amplitude density $\rho_s(x,t)$. The RSI transition amplitude density delocalizes between $t_i=0$ and $t=20$, then relocalizes between $t=20$ and $t_f=40$. The transition is smooth and continuous, with neither strange distortions, nor rapid oscillations, nor abrupt collapses. In particular, the mean position of the absolute value of $\rho_s(x,t)$ does not exhibit  zitterbewegung, as shown in Fig. 2. The small $(\pm 0.006)$ oscillations visible in $\rho_s(x,t)$ are present in both $\psi_+(x,t)$ and $\phi_+^\dagger(x,t)$, and decrease as the numerical accuracy is increased, suggesting they are due to numerical imprecision of the calculations. The negative energy parts of the same wavefunctions give the same results, but with the phase reversed. For the transition from Fig. 1(e) to Fig. 1(h), numerical calculations give $A_s = -0.607 -  0.161 i$ and $P_s=0.394$. Note that these differ from the CI values calculated earlier because the RSI wavefunctions are the positive energy projections of the CI  wavefunctions. If we use the same initial ($\psi_+$) and final ($\phi_+$) CI wavefunctions, the CI and RSI gives the same results for the transition amplitude and probability.

The RSI postulates that wavefunction and transition amplitude collapse never happens: upon an ideal (minimal disturbance) measurement at $t_f=40$, the RSI amplitude density does not change. After this measurement it continues to evolve (according to the DE) to the next ideal measurement. This is a generalization to all ideal measurements of the CI concept of eigenmeasurements. The case of non-ideal measurements is more complicated, and will be discussed in a later paper.
\section{Experimental Tests to Distinguish Between the CI and RSI} 
The CI and RSI of the DE can be experimentally distinguished by measuring whether a free spin-$\frac{1}{2}$ particle in vacuum displays zitterbewegung or not. This experiment is beyond the capabilities of current technology, but may be possible with future improvements in technology.

The CI of the DE and the conservation laws requires that the wavefunction of a single spin-$\frac{1}{2}$ particle must collapse instantaneously, in all reference frames. The RSI of the DE requires that the wavefunctions must evolve continuously and smoothly. An experimental test of whether instantaneous collapse occurs or not would distinguish between the CI and the RSI.

The CI predicts that the probability density at $t=20$ in the above experiment will be asymmetric in space, as shown in Fig. 1(b). The RSI predicts that the transition amplitude density at $t=20$ will be symmetric in space, as shown in Fig. 1(f). \textquotedblleft Weak measurement" experiments should be able to distinguish between these two predictions \cite{Struppa}. 
\section{Discussion} 
The KGE for spin-$0$ particles was discovered before the DE. The CI of the KGE predicts both positive and negative probability densities \cite{Greiner}. For these reasons, the KGE was believed to be wrong for many years. Then in 1934, Pauli and Weisskopf proposed a reinterpretation of the KGE: the probability densities could be multiplied by the particle's charge and reinterpreted as charge densities \cite{Pauli}. The existence of spin-$0$ particles with positive and negative charges was taken as a confirmation of this interpretation. But this makes the CI interpretations of the DE and the KGE inconsistent: why does an electron have a probability density, but a Higgs boson not have a probability density? In contrast, the RSI has a consistent interpretation of both the DE and the KGE: they both predict only transition amplitude densities, which are generally complex. Neither the DE transition amplitude density nor the KGE transition amplitude density needs to be multiplied by charge to give correct predictions for transition probabilities. 

The RSI predicts no zitterbewegung for a free spin-$\frac{1}{2}$ particle in vacuum. How can this be reconciled with the experimental simulation of zitterbewegung for such a particle \cite{Gerritsma2}? This simulation was done with a calcium ion in a Paul trap, and was actually a simulation of the CI of the DE, which is known to predict zitterbewegung. An experimental simulation of the RSI of the DE would not show zitterbewegung. How can the RSI be reconciled with the direct observation of zitterbewegung in solid state systems? For an electron in a crystal, the so-called zitterbewegung is actually the oscillatory motion of the electron traveling through a periodic lattice, and is a completely different phenomenon than the zitterbewegung of a free spin-$\frac{1}{2}$ particle in vacuum \cite{Zawadzki2}. 

Finally, the RSI has now given consistent interpretations of the Schr\"odinger equation \cite{Heaney3}, the KGE \cite{Heaney1}, and the DE. These new interpretations have resolved some of the puzzling aspects of the CI of these equations. These results suggest applying the RSI to resolve other puzzling aspects of quantum mechanics. Future papers will explore this possibility. 
 
\end{document}